\documentclass[
 aps, %prl,
 superscriptaddress,
 reprint,
 twocolumn,
 showpacs,preprintnumbers,
 amsmath,amssymb,
 pre,
]{revtex4}

\usepackage{graphicx} % Include figure files
\usepackage{dcolumn} % Align table columns on decimal point
\usepackage{bm} % bold math
\usepackage{color} % color
\usepackage{array,multirow}

\begin{document}

%\preprint{APS/123-QED}

\title{Network analysis of 3D complex plasma clusters in a rotating electric field}
\author{I.~Laut}
\email{laut@mpe.mpg.de}
\affiliation{Max-Planck-Institut f\"{u}r Extraterrestrische Physik, D-85741 Garching, Germany }
\author{C.~R\"{a}th}
\affiliation{Max-Planck-Institut f\"{u}r Extraterrestrische Physik, D-85741 Garching, Germany }
\author{L.~W\"{o}rner}
\affiliation{Max-Planck-Institut f\"{u}r Extraterrestrische Physik, D-85741 Garching, Germany }
\author{V.~Nosenko}
\affiliation{Max-Planck-Institut f\"{u}r Extraterrestrische Physik, D-85741 Garching, Germany }
\author{S.~K.~Zhdanov}
\affiliation{Max-Planck-Institut f\"{u}r Extraterrestrische Physik, D-85741 Garching, Germany }
\author{J.~Schablinski}
\affiliation{Christian-Albrechts Universit\"at zu Kiel, D-24118 Kiel, Germany }
\author{D.~Block}
\affiliation{Christian-Albrechts Universit\"at zu Kiel, D-24118 Kiel, Germany }
\author{H.~M.~Thomas}
\affiliation{Max-Planck-Institut f\"{u}r Extraterrestrische Physik, D-85741 Garching, Germany }
\author{G.~E.~Morfill}
\affiliation{Max-Planck-Institut f\"{u}r Extraterrestrische Physik, D-85741 Garching, Germany }
\date{\today}

\begin{abstract}
{Network analysis was used to study the structure and time evolution of driven three-dimensional complex plasma clusters. The clusters were created by suspending micron-size particles in a glass box placed on top of the rf electrode in a capacitively coupled discharge. The particles were highly charged and manipulated by an external electric field that had a constant magnitude and uniformly rotated in the horizontal plane. Depending on the frequency of the applied electric field, the clusters rotated in the direction of the electric field or remained stationary. The positions of all particles were measured using stereoscopic digital in-line holography. The network analysis revealed the interplay between two competing symmetries in the cluster. The rotating cluster was shown to be more cylindrical than the nonrotating cluster. The emergence of vertical strings of particles was also confirmed.
}
\end{abstract}

% PACS, the Physics and Astronomy

\maketitle
%\tableofcontents

\section{Introduction}
\label{sec:introduction}

Complex plasmas exist in various forms, from small clusters to large extended systems \cite{fortov2005, ivlev2012}. Two- and three-dimensional (3D) clusters are popular objects to study for two main reasons. First, they are used as model systems to study generic phenomena such as self-organization and transport, at the level of individual particles \cite{arp2004, totsuji2005, melzer2010, woerner2012, hyde2013}. Second, clusters can be used for diagnostic purposes, e.\,g. probing the plasma parameters at the position of particles \cite{arp2005, carstensen2010, nosenko2009}.

% External confinement.
The dust particles constituting a complex plasma cluster are highly charged and therefore have to be externally confined. For instance, in Ref.~\cite{arp2005} the particles were suspended in a short open glass tube placed on top of the rf electrode in a gas discharge. The authors suggested that the nearly isotropic particle cluster confinement was mainly determined by gravitational, electric and thermophoretic forces. A contribution from the ion drag force due to streaming ions was not observed. Dust particles suspended in a glass box can exhibit various structures, such as isolated single linear chains~\cite{kong2011}, vertical strings~\cite{woerner2012}, zigzag structures~\cite{melzer2006}, helical structures~\cite{tsytovich2007, hyde2013}, and Coulomb clusters with onionlike shells~\cite{arp2005, totsuji2005, ivanov2009}.

% 3D complex plasma crystals and clusters
Since the first observation of crystalline structure in 3D complex plasmas \cite{pieper1996dispersion}, the structural properties of such systems (consisting of about $10^4$ particles) have been examined thoroughly. For instance, domains of different lattice geometries were found\,\cite{zuzic2000, klumov2009, klumov2010}. Complex plasma clusters containing up to a hundred particles allowed an analysis at the level of individual particles and were exploited to study, e.g., the interaction force between the particles \cite{antonova2006}.

% Driven clusters.
Driving particle clusters with external forces adds an extra degree of flexibility to the experiments. One way of manipulating a particle cluster is the ``rotating wall'' technique, where an external rotating electric field is applied to the cluster \cite{nosenko2009, woerner2012}. In Ref.~\cite{woerner2012}, 3D particle clusters had a spheroidal shape, yet a competing cylindrical symmetry was also present. The latter is caused by streaming ions \cite{lampe2000, lampe2005}, which create an anisotropy in the interparticle potential and lead to vertical particle strings. To properly study this complicated structure, new analysis methods are clearly required.

% Tool: Network analysis
An advanced method of studying large systems is (complex) network analysis. Originating from ``classical'' networks such as power grids~\cite{watts1998} or the World Wide Web~\cite{albert1999}, complex network analysis~\cite{albert2002} has been adopted for a wide range of systems. Common to all these applications is the procedure of associating the constituents of a system (current generators or routers) with the nodes of a network, and their interactions (transmission lines between generators or connections between routers) with edges connecting them. While one strength of the network analysis is without doubt the access to vast and intricate objects~\cite{barabasi2011}, one other possible application is to small systems: By keeping track of all individual interactions throughout the analysis procedure, this approach may remain applicable where many other tools relying on some kind of averaging fail because of too weak statistics.

% In this paper ...
In this paper, we apply network analysis to driven 3D complex plasma clusters that were observed in Ref.~\cite{woerner2012}. In these spheroidal clusters, vertical strings were identified by introducing a certain fixed threshold (about a plasma screening length) to their transverse extent. This simple approach has its evident advantages but also limitations such as erroneously including passing by particles and a somewhat arbitrary threshold. With the help of multislice networks~\cite{mucha2010}, we now find strings in a natural way, and resolve them throughout the whole time series. Furthermore, the global structure of the clusters is analyzed in great detail by comparing network measures of the experimental data with null models. As we demonstrate in this paper, the network analysis is a powerful tool to analyze the structure of complex plasma clusters.

% The paper is organized as follows ..
The paper is organized as follows. In Sec.~\ref{sec:dynamically_driven_clusters} we briefly describe dynamically driven clusters. In Sec.~\ref{sec:networks_derived_from_distance_matrices} we examine particle strings of the clusters and their global geometry with the aid of network analysis. In Sec.~\ref{sec:confinement} we examine the particle confinement in the clusters by assuming a dynamical force balance. Finally, in Sec.~\ref{sec:summary}, we conclude with a summary of our results.

\section{Dynamically Driven Clusters}
\label{sec:dynamically_driven_clusters}

Particle clusters which we analyze in this paper were observed in the course of experiment of Ref.~\cite{woerner2012}. Below, we briefly outline the experimental procedure used in Ref.~\cite{woerner2012}.

\begin{figure}%%%%%%%%%%%%%%%%%%%%%%%%%%%%%%%%%%%%%%%%%%%%%%%%%%%%%%%%%%%%%%%%%
%\centering
\includegraphics[width=\columnwidth]{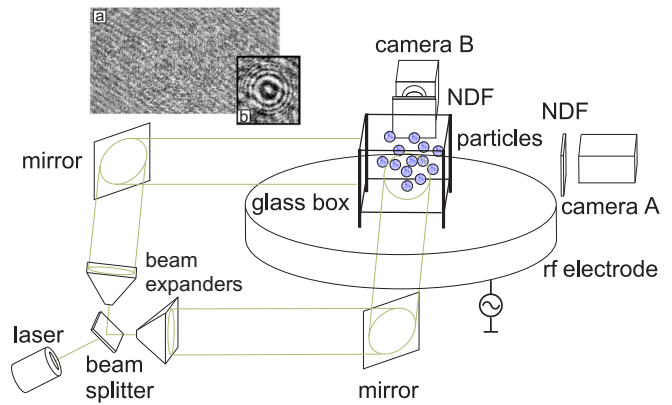}
\caption{Experimental setup. A 3D cluster of micron-size particles is suspended in the glass box mounted on top of the rf electrode in a capacitively coupled discharge. The particles are charged negatively and are manipulated by applying voltages on the conductive side plates of the box. By shifting sinusoidal signals on the adjacent plates by $\pi/2$, a rotating electric field is created at the position of particles. 
Expanded laser beams at $532$\,nm illuminate the particles from two perpendicular directions. The diffracted light is attenuated by neutral density filters (NDF) and registered by CCD cameras A and B. Insets show the interference patterns (a) of the whole cluster and (b) of an individual dust particle.}
\label{fig:1}
\end{figure}

% Experiment.
3D clusters of micron-size particles were suspended in a glass box mounted on top of the rf electrode in a capacitively coupled discharge in argon (see Fig.~\ref{fig:1}). Sinusoidal voltages were individually applied to the sides of the box, which are coated with indium tin oxide and are therefore conductive yet transparent. The phase shift between neighboring sides was set to $\pi/2$, which resulted in the electric field at the position of the particles that had a constant magnitude and rotated in the horizontal plane.

% Digital holography.
The particle coordinates were measured using a 3D imaging method, stereoscopic digital in-line holography \cite{kroll2008}. In two identical channels, expanded laser beams illuminate the particle cluster from two perpendicular directions. The two channels use the same $532$-nm laser with its output beam split in two parts. The diffracted light is registered directly by two CCD cameras operating at $50$ frames per second over a time interval of $10$\,s. In the resulting images, each particle is represented by a system of concentric circles (see insets in Fig.~\ref{fig:1}). The depth information is encoded in the intercircle spacing.

% Fig1 %%%%%%%%%%%%%%%%%%%%%%%%%%%%%%%%%%%%%%%%%%%%%%%%%%%%%%%%%%%%%%%%%%%%%%%%
%%%%%%%%%%%%%%%%%%%%%%%%%%%%%%%%%%%%%%%%%%%%%%%%%%%%%%%%%%%%%%%%%%%
%%%%%%%%%%%%%%%%%%%%%%%%%%%%%%%%%%%%%%%%%%%%%%%%%%%%%%%%%%%%%%%%%%%%%%%%%%%%%%%

% Cluster rotation.
Depending on the frequency of the applied electric field, the clusters rotated in the direction of the electric field (albeit with much lower frequency) or remained stationary. Two mechanisms of cluster rotation were proposed in Ref.~\cite{nosenko2009}. One mechanism is based on the interaction of the induced dipole moment of the cluster with the applied field, the other on the ion drag force. In the present paper, we analyze a rotating cluster driven at $5$\,kHz and a stationary one driven at $1$\,kHz.

% Cluster compression.
Whether rotating or not, the clusters were significantly compressed in the radial direction by the applied field. The specific mechanism of this compression is not clear. The ponderomotive force \cite{lamb1983}, a naturally expected candidate, is evidently of minor importance in the conditions of this experiment. Indeed, the clusters driven at the frequencies of $1$ and $5$\,kHz were actually of the same size, see Fig.~\ref{fig:8}, whereas the ponderomotive force has an inverse quadratic dependence on frequency.

% Strings, structure.
The main result of Ref.~\cite{woerner2012} is the formation of vertical particle strings in driven clusters. A comparison with dust dynamics simulations \cite{ludwig2012}, where the vertical ordering parameter depended on the thermal Mach number of flowing ions, showed reasonable agreement.

% Competing symmetries.
Structural changes in the global cluster symmetry were also observed in Ref.~\cite{woerner2012}. Driven clusters had elements of competing symmetries that changed from one type to the other while the cluster rotated. This topic was not pursued further in part for the lack of proper analysis technique.

\section{Networks derived from Distance Matrices}
\label{sec:networks_derived_from_distance_matrices}

\subsection{Characterizing Particle Clusters}
\label{subsec:characterizing_particle_clusters}
% analyze ts with networks
In order to analyze the cluster structure, we adopted a network approach similar to those which have recently been successfully applied to (nonlinear) time series~\cite{donner2010, marwan2009}. There, a scalar time series of~$T$ steps is embedded into an $m$-dimensional phase space. The corresponding ${(T-m)\times(T-m)}$ recurrence matrix~$R_{ij}$ has unity entries if the distance between the~$i$th and~$j$th point in phase space is below a certain threshold, and zero otherwise. Interpreting the recurrence matrix as the adjacency matrix of a network, one can investigate the time series by means of network theory. The resulting network is called unweighted, as all connections (links) between the network nodes have the same strength.

% now: spatial distance -> A
We will adopt this approach in Sec. \ref{subsec:global_structure} to determine the global structure of clusters: The ${n\times n}$ adjacency matrix then connects two particles whose difference in cylindrical radius is below a predefined threshold, where~$n$ is the number of particles.

% also: communities in weighted networks
We will also use a modified approach, where the components of the adjacency matrix are different from zero or one. For such a weighted network, we will consider communities, i.e., sets of nodes which are more connected to each other than to the rest of the network, in order to find vertical strings in the cluster.

\subsection{Detection of Strings}
\label{subsec:detection_of_strings}
% community detection with modularity
Before we focus on the global structure of the clusters, we briefly outline the possibility for string detection using community assignment~\cite{girvan2002} in networks. A popular method to quantify communities is by the quality function \textit{modularity} that compares the number of intracommunity edges to the expectation value of a network with randomized links~\cite{newman2004finding, newman2006finding}.

% multislice modularity
With the tool of multislice networks~\cite{mucha2010, bassett2013} it is possible to find communities in systems that evolve over time.
In addition to adjacency matrices $A_{ijt}$ for each time step~$t$ there is an interslice coupling parameter $C_{jtr}$ connecting node $j$ at time $t$ to itself at time $r$. The \textit{multislice modularity}~\cite{mucha2010} then reads
 \begin{equation} \label{equ:quality_function}
Q \propto \sum\limits_{ijtr} \left[ \left( A_{ijt} - \gamma \frac{k_{it} k_{jt}}{2 m_t} \right) \delta_{tr} + \delta_{ij} C_{jtr} \right] \delta(g_{it}, g_{jr}),
 \end{equation}
where~$\gamma$ is a free resolution parameter, ${k_{jt} = \sum_{i} A_{jit}}$ the strength of an individual node, ${m_t = \sum_{j} k_{jt}}$,  and $g_{it}$ is the community assignment of particle~$i$ at time~$t$.

% slices
We choose the \textit{slices} $A_{ijt}$ to be proportional to the element-wise inverse distance matrix of the projection of the particles on the $xy$ plane at a given time step,
 \begin{equation} \label{equ:Aijt}
A_{ijt} = 
\begin{cases}
        \Delta/d_{ij} & \text{for } i \neq j \\
        0                       & \text{otherwise}.
\end{cases}
 \end{equation}
Here, $d_{ij} = \sqrt{(x_i - x_j)^2 + (y_i - y_j)^2}$ is the horizontal distance between two particles at time $t$ and $\Delta$ the minimal horizontal distance in the time series such that~$A_{ijt}$ is at most one. The link between two particles is thus stronger, the nearer they are to each other, and one can consider the communities of the resulting network as strings of the cluster.

% interslice coupling
In order to resolve the communities in time, the components of the multislice adjacency matrix connecting a particle with itself at the consecutive time step were set to ${C_{jtr} = 0.1 \delta_{t, r+1}}$.

% algorithm
The algorithm used~\cite{mucha2010} is an adaptation of the Louvain method~\cite{blondel2008} that aggregates nodes locally to small communities which are the nodes of a new network at a later iteration step until~$Q$ is maximized.

% Fig2 %%%%%%%%%%%%%%%%%%%%%%%%%%%%%%%%%%%%%%%%%%%%%%%%%%%%%%%%%%%%%%%%%%%%%%%%
 \begin{figure}%%%%%%%%%%%%%%%%%%%%%%%%%%%%%%%%%%%%%%%%%%%%%%%%%%%%%%%%%%%%%%%%
%\centering
 \includegraphics[width=\columnwidth]{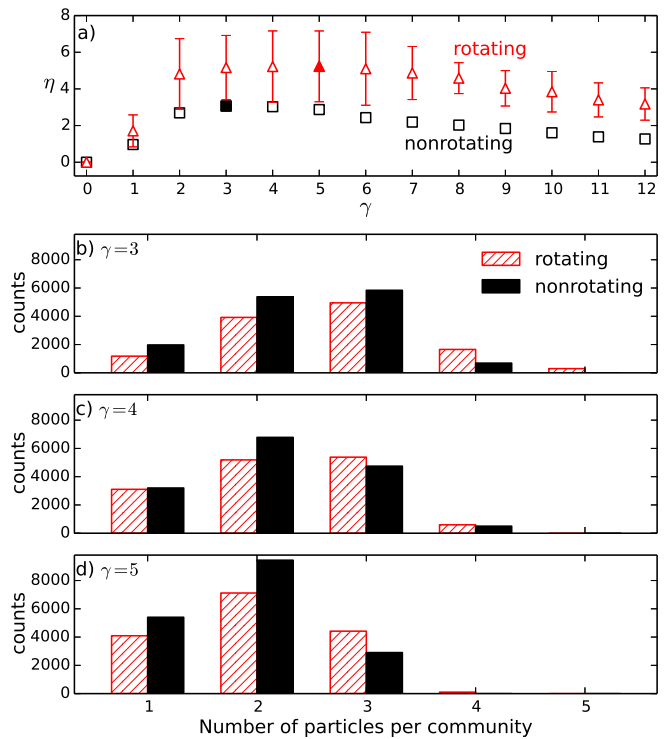}%
 \caption{Dependency of detected communities on resolution parameter~$\gamma$.
(a) Ratio~$\eta$ of intercommunity distance to community size for different values of~$\gamma$ for the rotating (triangles) and nonrotating (squares) cluster. The maximum values are plotted as solid symbols. The standard deviations are plotted as error bars and are concealed by the symbols in the case of the nonrotating cluster.
(b)--(d) Histograms of the number of particles per community for different values of~$\gamma$.}
 \label{fig:2}
 \end{figure}%%%%%%%%%%%%%%%%%%%%%%%%%%%%%%%%%%%%%%%%%%%%%%%%%%%%%%%%%%%%%%%%%%
%%%%%%%%%%%%%%%%%%%%%%%%%%%%%%%%%%%%%%%%%%%%%%%%%%%%%%%%%%%%%%%%%%%%%%%%%%%%%%%

% Fig3 %%%%%%%%%%%%%%%%%%%%%%%%%%%%%%%%%%%%%%%%%%%%%%%%%%%%%%%%%%%%%%%%%%%%%%%%
 \begin{figure}%%%%%%%%%%%%%%%%%%%%%%%%%%%%%%%%%%%%%%%%%%%%%%%%%%%%%%%%%%%%%%%%
 \includegraphics[width=\columnwidth]{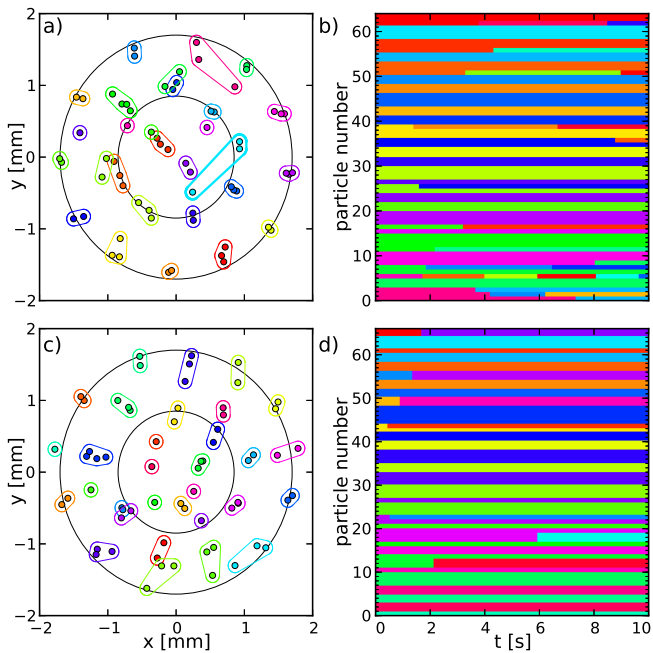}%
 \caption{Time evolution of detected strings for resolution parameter $\gamma = 4$.
(a)~Projection of the rotating cluster on the $xy$ plane at $t=0$. Particles which were found to be in the same string  are grouped together and have the same color. (b)~String affiliation of particles over time. A transition from one string to another appears as a change in color of the line corresponding to the particle. (c),(d) Same for the nonrotating cluster.}
 \label{fig:3}
 \end{figure}%%%%%%%%%%%%%%%%%%%%%%%%%%%%%%%%%%%%%%%%%%%%%%%%%%%%%%%%%%%%%%%%%%
%%%%%%%%%%%%%%%%%%%%%%%%%%%%%%%%%%%%%%%%%%%%%%%%%%%%%%%%%%%%%%%%%%%%%%%%%%%%%%%

% resolution parameter: define eta
The resolution parameter~$\gamma$ is considered optimal when the average extent of a community is small compared to the intercommunity distance, i.e., when
\begin{equation} \label{equ:gamma_quality}
\eta = \langle d_{i,\text{next}}^t / d_{i,\text{same}}^t \rangle
\end{equation}
is maximized. Here, $d_{i,\text{next}}^t$ is the horizontal distance of particle~$i$ to the next particle of another community at time~$t$, $d_{i,\text{same}}^t$ is the average horizontal distance to the particles in the same community if the particle is in a community with more than one particle, and infinity otherwise. The brackets denote the average over all particles and time steps. %$\langle \rangle$

% result: use gamma=4
As can be seen in Fig.~\ref{fig:2}\,(a), $\eta$ varies rather weakly in the range of resolution parameter~$\gamma$ from 2 to 6 around the maxima for the clockwise rotating and nonrotating clusters. The number of particles per community is very sensitive to~$\gamma$ [see Figs.~\ref{fig:2}\,(b)--(d)]. For the three values of $\gamma$ considered here, the algorithm finds more small communities of one or two particles in the case of the nonrotating cluster, while larger communities with at least three particles are more likely for the rotating  cluster.

% now: communities -> strings
By considering the communities as the vertical particle strings in the cluster, one thus obtains strings of different sizes depending on~$\gamma$, in contrast to a more traditional approach of grouping particles whose distance is below a certain value as in Ref.~\cite{woerner2012}.

% time evolution
The strings for~$\gamma=4$ and their evolution in time for the rotating and nonrotating clusters are shown in Fig.~\ref{fig:3}. It is evident that in the case of the nonrotating cluster there are fewer transitions between the strings. 

The strings are quite robust against encounters with only occasionally passing by particles: During their passage, these roaming particles are not considered to be part of the string. The cases where the horizontal distance between the particles is small, while a large vertical distance prohibits physical correlation, do not persist in time. These events are thus not considered as strings by the community-finding algorithm. Indeed, the mean vertical distance between particles of the same string is 0.79\,mm (rotating cluster) and 0.86\,mm (nonrotating cluster), and the events where this distance is larger than $1.2$\,mm are not frequent (less than 3\,\%) for both clusters. 
By inspection of Fig.~\ref{fig:3}\,(a) alone, it is not obvious why, for example, the particle near the center belongs to the highlighted longish community. In order to understand this, the whole time series has to be considered, as this roaming particle quickly joins the two remaining particles and forms a persistent string. This special feature of community assignment in networks may be a powerful tool within a wide range of possible applications.

% Fig4 %%%%%%%%%%%%%%%%%%%%%%%%%%%%%%%%%%%%%%%%%%%%%%%%%%%%%%%%%%%%%%%%%%%%%%%
\begin{figure}%%%%%%%%%%%%%%%%%%%%%%%%%%%%%%%%%%%%%%%%%%%%%%%%%%%%%%%%%%%%%%%%
\centering
\includegraphics[width=\columnwidth]{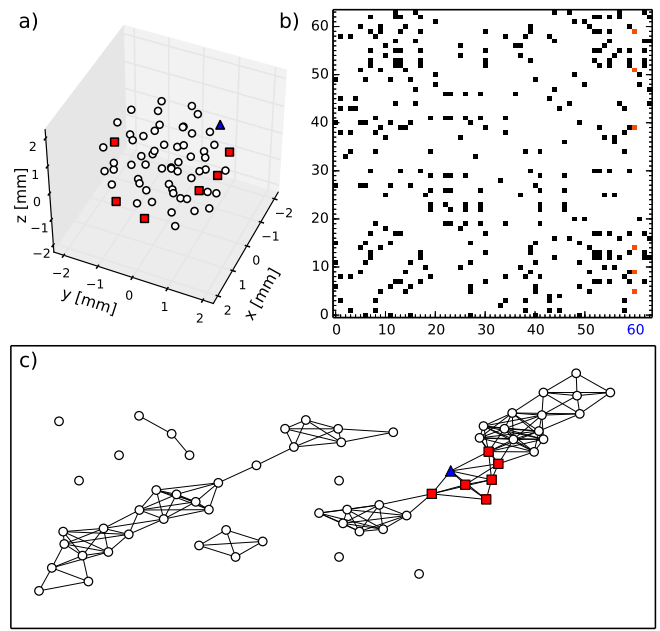}
\caption{Method for generating an unweighted network from the data.
(a) 3D plot of the clockwise rotating cluster.
The 60th particle is plotted as a blue triangle and the particles $j$ satisfying $\left| \rho_{60} - \rho_j  \right| < \epsilon$ are plotted as red squares.
(b) The corresponding adjacency matrix. The nodes connected to the 60th node can be read off the 60th column (or line) of the matrix.
(c) Representation of the network. The nodes representing the particles have the same markers as in (a).}
\label{fig:4}
\end{figure}%%%%%%%%%%%%%%%%%%%%%%%%%%%%%%%%%%%%%%%%%%%%%%%%%%%%%%%%%%%%%%%%%%%
%%%%%%%%%%%%%%%%%%%%%%%%%%%%%%%%%%%%%%%%%%%%%%%%%%%%%%%%%%%%%%%%%%%%%%%%%%%%%%%

\subsection{Global Structure of the Clusters}
\label{subsec:global_structure}
% outline: cluster -> unweighted network
The global structure of the clusters is analyzed by means of unweighted networks. The adjacency matrix connects two particles whose difference in cylindrical radii is small enough. The procedure is sketched in Fig.~\ref{fig:4}: In (a), the 60th particle of the cluster is plotted as a blue triangle and the particles with comparable cylindrical radii as red squares. The latter can be read off the 60th column of the corresponding adjacency matrix, (b). A representation of the resulting network can be seen in Fig.~\ref{fig:4}\,(c). At this time step, the network consists of two main components (groups of connected nodes), which correspond to the two ring regions of the cluster, and various smaller components.

% outline: ratio of "best" null model -> ratio of competing structures
The network thus obtained is analyzed using network measures and the results are compared to those of a network from a null model where a certain fraction of points is in perfectly spherical order, and the rest in cylindrical order. The ratio that shows the best agreement with the experimental data will be considered as the ratio of the competing spherical and cylindrical geometries of the cluster.

% define adj. matrix
We define the adjacency matrix as
\begin{equation}
\label{equ:adj_matrix_cyl}
A^{\text{cyl}}_{ij}(\epsilon) = \Theta\left(\epsilon - \left| \rho_i - \rho_j  \right|\right)- \delta_{ij},
\end{equation}
where $\Theta(\cdot)$ is the Heaviside function, ${\rho_{i} = \sqrt{x_i^2 + y_i^2}}$ is the cylindrical radius, and $\epsilon$ an appropriate threshold. The Kronecker $\delta$ sets the diagonal terms to zero in order to avoid self-loops.

% alpha = 0.1
The threshold $\epsilon$ was chosen to be a fraction $\alpha$ of the mean difference in cylindrical radius: ${\epsilon= \alpha \langle \left| \rho_i - \rho_j  \right| \rangle}$. The brackets denote the average over all particles and time steps. Throughout this study we used $\alpha=0.1$.

\begin{figure}%%%%%%%%%%%%%%%%%%%%%%%%%%%%%%%%%%%%%%%%%%%%%%%%%%%%%%%%%%%%%%%%
 \centering
 \includegraphics[width=\columnwidth]{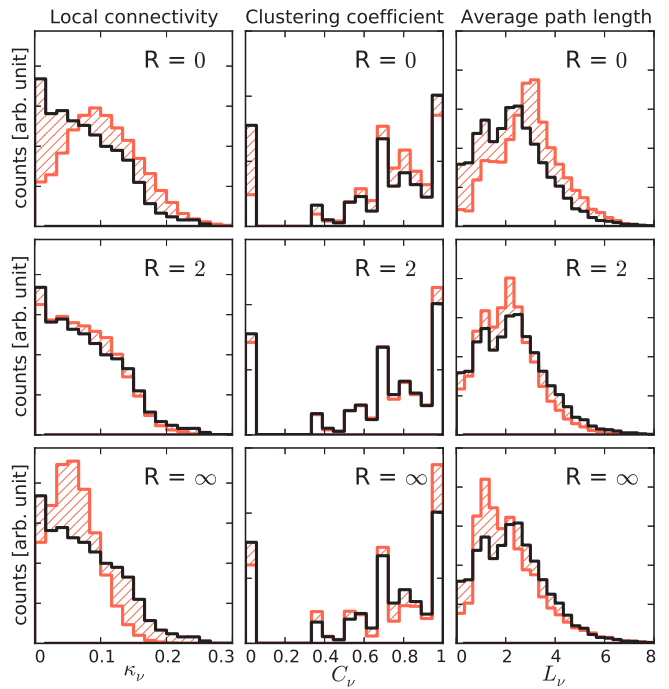}%
 \caption{Comparison of network measures for the clockwise rotating cluster with the measures for null models of different ratios $R$. The histograms (for all particles and time steps) of measures on networks of the clockwise rotating cluster are plotted in black. They are identical in all three rows. The null models are plotted in red (grey) with $R = 0$~(top row), $2$~(middle row), and $\infty$~(bottom row), where $R$ is the ratio of particles from a perfectly spherical geometry to particles from a cylindrical geometry. First column: local connectivity~$\kappa_\nu$, second column: clustering coefficient~$C_\nu$, third column: average path length~$L_\nu$. The differences are marked as hatched areas. The good agreement of all three network measures in the middle row is clearly visible.}
 \label{fig:5}
 \end{figure}

\begin{figure}%%%%%%%%%%%%%%%%%%%%%%%%%%%%%%%%%%%%%%%%%%%%%%%%%%%%%%%%%%%%%%%%
%\centering
 \includegraphics[width=\columnwidth]{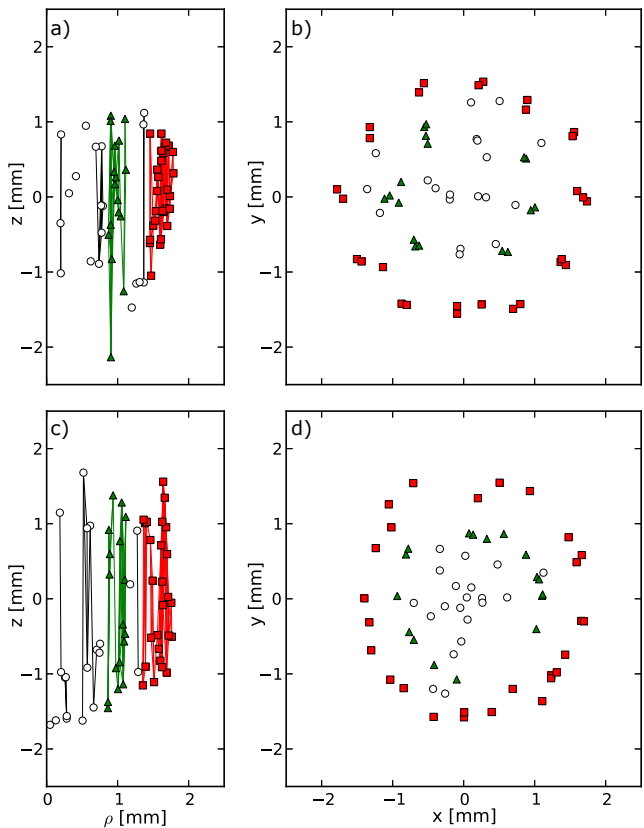}%
 \caption{Comparison of experimental cluster~(top) to null model~(bottom). Projections of the clockwise rotating cluster onto the $\rho z$ plane~(a) and $xy$ plane~(b). The two main components of $A^{\text{cyl}}$ are plotted as triangles (squares) and correspond to the inner (outer) ring regions of the cluster. For the sake of clarity the network edges are only plotted in panel (a). The network is identical to the one in Fig.~\ref{fig:4}, only the positions of the nodes are chosen differently: While in Fig.~\ref{fig:4} a more abstract representation of the network was chosen to show its structure, here the positions correspond to the projections of the particles. (c),(d)~The same for the null model whose network measures showed the best agreement with the cluster: The ratio of particles in a spherical arrangement to particles in a cylindrical arrangement is $R = 2$. }
 \label{fig:6}
 \end{figure}

% null model: mixture of two structures
The null models are artificial structures with a predefined ratio ${R=n^{\text{sph}}/n^{\text{cyl}}}$ of the number of particles in a perfect spherical structure to the number of particles in a perfect cylindrical structure. The total number of particles ${n=n^{\text{sph}}+n^{\text{cyl}}}$ of the null models is equal to the number of particles in the experimental data. We use two-shell null models with different ratios $R =  0, \,1/3, \,1/2, \,1, \,2, \,3, \,\infty$. Each model is constructed as follows. The cylindrical structure consists of two concentric cylinders with the same (cylindrical) radii as the main components of the experimental data ($\rho_1=1.0$\,mm, $\rho_2=1.6$\,mm). For a given $n^{\text{cyl}}$, the ratio of the particle number in the inner cylinder to that in the outer cylinder is chosen to be equal to $\rho_1/\rho_2$. The two shells of the spherical structure have (spherical) radii $r_1=1.1$\,mm and $r_2=1.7$\,mm. 
A uniformly distributed random noise of amplitude $0.15$\,mm is added to the positions of all particles. It compensates the simplification made by choosing the two-shell model---the clusters are apparently three-shell structured[see, e.g., Fig.~\ref{fig:4}\,(c)]. The presence of the innermost shell, consisting of only a few particles, does not change the results significantly, and the two-shell model, being properly adjusted, agrees well with the experiment (see Fig.~\ref{fig:5}, middle row). See Figs.~\ref{fig:6}\,(c) and (d) for the projections of such a null model with $R=2$ on the $\rho z$ and $xy$ plane.

% Fig5 %%%%%%%%%%%%%%%%%%%%%%%%%%%%%%%%%%%%%%%%%%%%%%%%%%%%%%%%%%%%%%%%%%%%%%%%
 %%%%%%%%%%%%%%%%%%%%%%%%%%%%%%%%%%%%%%%%%%%%%%%%%%%%%%%%%%%%%%%%%%
%%%%%%%%%%%%%%%%%%%%%%%%%%%%%%%%%%%%%%%%%%%%%%%%%%%%%%%%%%%%%%%%%%%%%%%%%%%%%%%

% degree centrality
As a first network measure, the degree centrality~$k_\nu$ counts the number of nodes that are connected to node~$\nu$. It is defined as \cite{donner2010}
\begin{equation}
\label{equ:degree_centrality}
k_\nu =  \sum\limits_{i=1}^n A_{\nu, i}.
\end{equation}
Normalizing it by the maximum possible value yields the local connectivity ${\kappa_{\nu} = k_{\nu}/ (n-1)}$. Taking into account only the immediate neighbors of the node \footnote{Note that depending on the definition of the adjacency matrix, connected nodes (neighbors) of the network do not necessarily represent particles that have a small spatial separation.}, this measure may be regarded as the local particle density of the cluster. 

% clustering coefficient
The clustering coefficient~$C_\nu$ \cite{donner2010} acts on intermediate scales. It evaluates the number $N_\nu^\Delta$ of links between neighbors of a given node versus the maximum possible number $k_\nu(k_\nu-1)/2$:
\begin{equation}
\label{equ:clustering}
C_\nu = \frac{2}{k_\nu(k_\nu-1)} N_\nu^\Delta = \frac{1}{k_\nu(k_\nu-1)} \sum\limits_{i, j=1}^n A_{\nu, i} A_{i, j} A_{j, \nu}
\end{equation}

% avg path
Finally, the average path length \cite{donner2010} is considered. For every node $\nu$ of the network, the minimum numbers $l_{\nu, i}$ of edges that have to be traversed to get to any other node $i$ of the same component is calculated. The average path length $L_\nu$ is then calculated by averaging over all nodes~$i$ that are in the same component as $\nu$:
\begin{equation}
\label{equ:average_path}
L_{\nu} = \langle l_{\nu, i} \rangle
\end{equation}

% chose best R
For each time step of the experimental data, the corresponding networks of the data and of a null model with a given ratio $R$ are created and analyzed with the network measures at hand. The histograms of these measures for all particles and all time steps of the data are compared, and the ratio $R$ of the null model with  the best agreement is considered to be the ratio of spherical to cylindrical order of the cluster.

% example result
In Fig.~\ref{fig:5}, the results for $A^{\text{cyl}}$ of the clockwise rotating cluster are shown and compared to the null models with ratios ${R = 0}$ (purely cylindrical structure, top), 2 (smallest deviation, middle) and $\infty$ (purely spherical structure, bottom). Given the good agreement for $R= 2$,  one can argue that the spherical geometry of the cluster is two times more pronounced than the cylindrical geometry. The projections of the null model with ${R = 2}$ on the $\rho z$ and $xy$ planes are plotted in Fig.~\ref{fig:6} (bottom) and compared to the clockwise rotating cluster (top).

% Fig6 %%%%%%%%%%%%%%%%%%%%%%%%%%%%%%%%%%%%%%%%%%%%%%%%%%%%%%%%%%%%%%%%%%%%%%%%
 %%%%%%%%%%%%%%%%%%%%%%%%%%%%%%%%%%%%%%%%%%%%%%%%%%%%%%%%%%%%%%%%%%
%%%%%%%%%%%%%%%%%%%%%%%%%%%%%%%%%%%%%%%%%%%%%%%%%%%%%%%%%%%%%%%%%%%%%%%%%%%%%%%

% A sph
In order to test our approach, the analysis has been repeated, this time comparing the spherical radii of the particles. The adjacency matrix now reads
\begin{equation}
\label{equ:adj_matrix_sph}
A^{\text{sph}}_{ij}(\epsilon) = \Theta\left(\epsilon - | r_i - r_j |\right) - \delta_{ij}
\end{equation}
in the same notation as in Eq.~\ref{equ:adj_matrix_cyl} where $r$ is now the spherical radius and ${\epsilon= \alpha \langle \left| r_i - r_j  \right| \rangle}$ with $\alpha=0.1$.

The results of the analysis of the rotating and nonrotating clusters for both adjacency matrices are summarized in Table~\ref{table:1}. Even though the results for the cluster geometry are not the same for $A^{\text{cyl}}$ and $A^{\text{sph}}$, the general observation remains the same: One finds comparable results in the cases of clockwise and counterclockwise rotation, while the value of~$R$ is greater for the nonrotating cluster.

\begin{figure}%%%%%%%%%%%%%%%%%%%%%%%%%%%%%%%%%%%%%%%%%%%%%%%%%%%%%%%%%%%%%%%%
%\centering
 \includegraphics[width=\columnwidth]{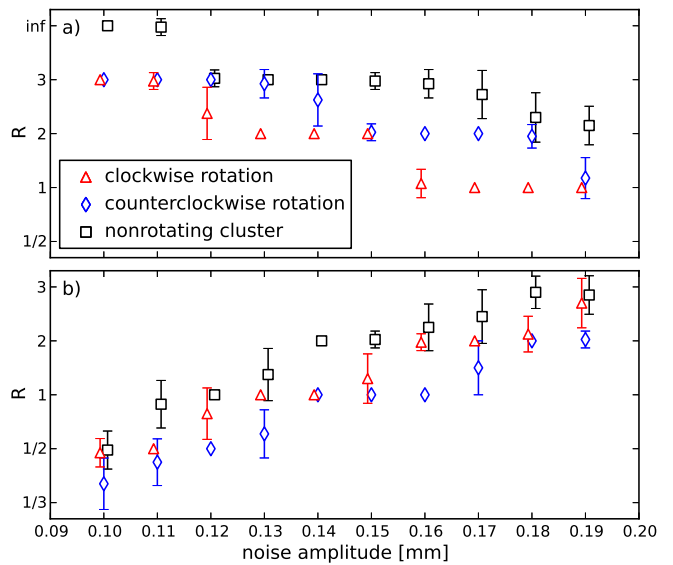}%
 \caption{Ratio $R = n^{\text{sph}}/n^{\text{cyl}}$ vs. noise amplitude of null models for $A^{\text{cyl}}$~(a) and $A^{\text{sph}}$~(b). The results are averaged over 40 calculations, and the standard deviations are plotted as error bars. The results in Table~\ref{table:1} correspond to a noise amplitude of $0.15$\,mm.}
 \label{fig:7}
 \end{figure}

% noise
The dependence of the best ratio~$R$ on the noise amplitude of the null models is shown in Fig.~\ref{fig:7}. Noticeably, decreasing the noise amplitude generally shifts the optimal ratios~$R$ to higher values for $A^{\text{cyl}}$ but towards lower values for $A^{\text{sph}}$. This can be understood as in the first case the difference in the cylindrical radii mostly determines whether nodes are connected or not. Hence, in order to find the best agreement, more particles of spherical structure have to be added when the noise level is reduced, increasing~$n^{\text{sph}}$ and thus increasing $R = n^{\text{sph}}/n^{\text{cyl}}$. In the second case of $A^{\text{sph}}$, decreasing  the noise amplitude is to be compensated by adding more cylindrically structured particles, yielding a lower value for~$R$. 

% fig7  %%%%%%%%%%%%%%%%%%%%%%%%%%%%%%%%%%%%%%%%%%%%%%%%%%%%%%%%%%%%%%%%%%%%%%%
 %%%%%%%%%%%%%%%%%%%%%%%%%%%%%%%%%%%%%%%%%%%%%%%%%%%%%%%%%%%%%%%%%%
%%%%%%%%%%%%%%%%%%%%%%%%%%%%%%%%%%%%%%%%%%%%%%%%%%%%%%%%%%%%%%%%%%%%%%%%%%%%%%%

% CONCLUSION
The results depend rather sensitively on the  noise amplitude of the null models (see Fig.~\ref{fig:7}). The exact values of the ratios~$R$ have thus to be treated with caution, but they are a convenient tool for comparing structures. Considering Fig.~\ref{fig:7}, it is not possible to name the exact value of the ratio of the competing spherical and cylindrical geometries of the clusters, but it is evident that the nonrotating cluster systematically shows a higher value of~$R$. The geometry is thus more spherical in the case of the nonrotating cluster, while the rotating clusters have a more pronounced cylindrical geometry.

\begin{figure}%%%%%%%%%%%%%%%%%%%%%%%%%%%%%%%%%%%%%%%%%%%%%%%%%%%%%%%%%%%%%%%%
%\centering
 \includegraphics[width=\columnwidth]{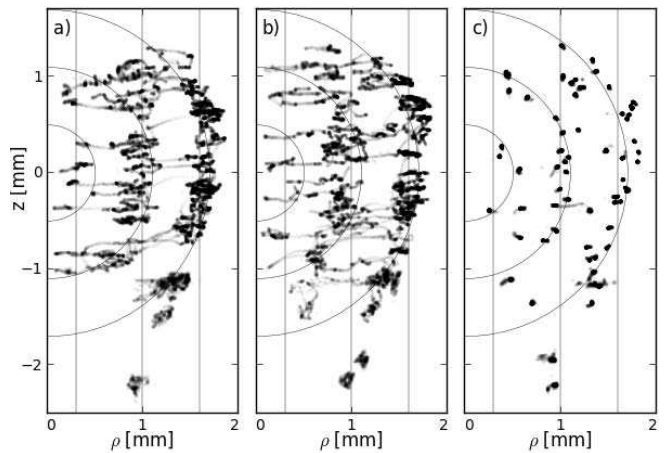}%
 \caption{Projection of particle positions on the $\rho z$\,plane for clockwise~(a) and counterclockwise~(b) rotating, and nonrotating~(c) clusters. Vertical lines and semicircles are adjusted to the cylindrical structure of the rotating clusters and to the spherical structure of the nonrotating clusters, respectively, in order to guide the eye. Particle positions at each time step correspond to a transparent marker, leading to lines in the event of a string transition and solid structures if the particle stays in the same position throughout the time series.}
 \label{fig:8}
 \end{figure}

% global structure as schella did: hard to see
Another possibility to examine the structure of a cluster is to merely plot the particle $z$ positions vs. cylindrical radius $\rho$~\cite{arp2004, totsuji2005, schella2013} (see Fig.~\ref{fig:8}). In this projection, spheres appear as semicircles and cylindrical structures as vertical lines. Comparing the projections in Fig.~\ref{fig:8}, one can see that in the case of rotation [Figs. \ref{fig:8}(a) and (b)] the particles at the bottom and the top of the cluster have moved outward, while the particles at $z \simeq 0$ appear to have moved inward compared to the nonrotating cluster, Fig. \ref{fig:8}(c). The rotating clusters thus appear more cylindrical, while the spheres are more pronounced in the nonrotating case, but there is neither strong visual evidence nor a numerical result as in the case of network analysis.

% table 1 %%%%%%%%%%%%%%%%%%%%%%%%%%%%%%%%%%%%%%%%%%%%%%%%%%%%%%%%%%%%%%%%%%%%%
 \begin{table}%%%%%%%%%%%%%%%%%%%%%%%%%%%%%%%%%%%%%%%%%%%%%%%%%%%%%%%%%%%%%%
 \caption{
%Comparison of results from network analysis with confinement parameter from estimated 'Yukawa force'.
Ratio $R = n^{\text{sph}}/n^{\text{cyl}}$ of the competing geometries (spherical and cylindrical) and anisotropic confinement parameters $\Omega_{\rho, z}$ of the clockwise (CW) and counterclockwise (CCW) rotating, and nonrotating clusters. The values of $R$ shown here provided the best agreement for network measures on the cylindrical~($A^{\text{cyl}}$) and spherical~($A^{\text{sph}}$) properties of the clusters. The results were averaged over 40 calculations, with the standard deviation indicated as error. The confinement parameters $\Omega_{\rho, z}$ [Eq.~(\ref{eq:confinement_frequency})] were calculated from the linear fitting parameters of the projections of the Yukawa forces (see Fig.~\ref{fig:9}). } 
 \begin{flushright}
 \begin{tabular}{m{2.3cm}            m{1.9cm}       m{1.9cm}       m{1.9cm} }
    \hline 
    \hline 
    \\[-2.0ex]  % seen it here: http://everythingyouforgetaboutlatex.blogspot.de/
                                   &   \multicolumn{3}{c}{Cluster rotation}     \\[0.3ex]
    \cline{2-4}
    \\[-2.0ex]
    \multicolumn{1}{l}{Parameter}  & CW           & CCW          & Nonrotating  \\[0.3ex]
    \hline
    \\[-2.0ex]
    $R$ for $A^{\text{cyl}}$       & $2.0\pm0.2$  & $2.0\pm0.2$  & $3.0\pm0.2$  \\
    $R$ for $A^{\text{sph}}$       & $1.3\pm0.5$  & $1.0\pm0.2$  & $2.0\pm0.2$  \\[0.3ex]
    \hline
    \\[-2.0ex]
    $\Omega_\rho$ [s$^{-1}$]       & $19\pm2$      & $12\pm15$   & $16.2\pm0.8$ \\
    $\Omega_z$ [s$^{-1}$]          & $33\pm1$      & $36\pm2$    & $32.0\pm0.3$ \\[0.3ex]
    \hline 
    \hline
 \end{tabular}
 \end{flushright}
 \label{table:1}
 \end{table}%%%%%%%%%%%%%%%%%%%%%%%%%%%%%%%%%%%%%%%%%%%%%%%%%%%%%%%%%%%%%%%%%%%
%%%%%%%%%%%%%%%%%%%%%%%%%%%%%%%%%%%%%%%%%%%%%%%%%%%%%%%%%%%%%%%%%%%%%%%%%%%%%%%

\section{Confinement Anisotropy}
\label{sec:confinement}
% force decomposition
The difference in structure between rotating and nonrotating clusters stems from the particulars of the mutual particle interactions and the forces confining the cluster. Even though it is a rather difficult task to explore the forces controlling the cluster dynamics in detail, it is instructive to assume that globally the system is in quasiequilibrium. The latter is determined by the balance of repulsion $\mathbf{F}^{\text{rep}}$ via the Yukawa forces, neutral gas friction  $\mathbf{F}^{\text{fr}}$, inertial forces $\mathbf{F}^{\text{in}}$, and confinement $\mathbf{F}^{\text{conf}}$ provided by all other forces \footnote{The confinement is provided by gravity, electric field of the plasma sheath, rotating electric field, ion drag force due to streaming ions, and ion wake mediated interaction.}:
\begin{equation}
\label{eq:balance}
\langle \mathbf{F}^{\text{rep}} \rangle +
\langle \mathbf{F}^{\text{fr}} \rangle +
\langle \mathbf{F}^{\text{in}} \rangle +
\langle \mathbf{F}^{\text{conf}} \rangle = 0.
\end{equation}
Here, the stochastic averaging is assumed to be performed. 

\begin{figure}%%%%%%%%%%%%%%%%%%%%%%%%%%%%%%%%%%%%%%%%%%%%%%%%%%%%%%%%%%%%%%%%
%\centering
 \includegraphics[width=\columnwidth]{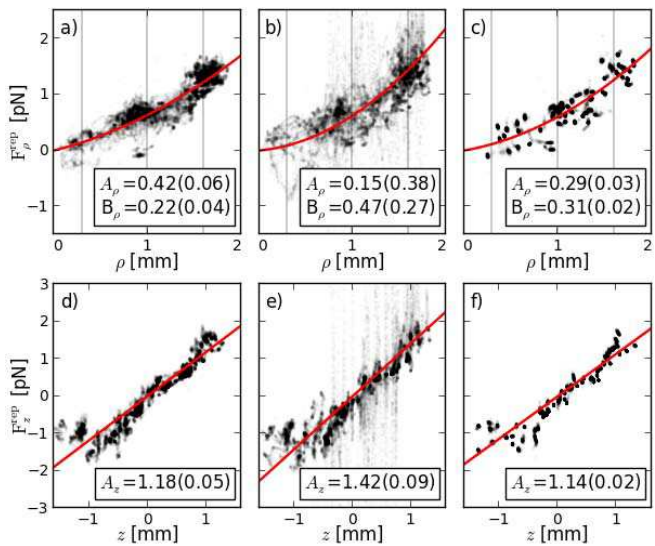}%
 \caption{Radial profiles of the repulsive Yukawa-type interactions of the particles for clockwise~(a) and counterclockwise~(b) rotating, and  nonrotating~(c) clusters. The cylindrical projection of the Yukawa forces is plotted vs. the cylindrical radius $\rho$. The solid lines are the best-rms-fit parabolas [Eq. (\ref{eq:fit})] to the force profiles. The insets show the mean values of $A_\rho$ in pN/mm and $B_\rho$ in pN/mm$^2$ which were calculated for each time step, with the standard deviations indicated as errors. The vertical lines are the same as in Fig.~\ref{fig:8}. (d)--(f) The same for the $z$ profiles of the Yukawa forces, here, the few particles suspended below the cluster at $z\simeq-2$\,mm (see Fig.~\ref{fig:8}) were left out. The insets show the parameter $A_z$ in pN/mm of the linear fit to the profiles. }
 \label{fig:9}
 \end{figure}

% Yukawa
The Yukawa forces, governing the mutual repulsion between the particles, can be directly calculated from the data:
\begin{equation}
\label{eq:yukawa}
\mathbf{F}_i^{\text{rep}} = 
-\left(Ze\right)^2 \nabla_{\mathbf{r}_i} \sum_{j\neq i}^N 
\frac{ \exp \left( -\left|\mathbf{r}_i-\mathbf{r}_j\right| / \lambda \right) }{ \left|\mathbf{r}_i-\mathbf{r}_j\right| },
\end{equation} % ~R_{i,j}=\left|\mathbf{r}_i-\mathbf{r}_j\right|
where $\mathbf{r}_i$ is the position of particle~$i$, $Ze=50000e$ is the particle charge ($e$ is the elementary charge) and $\lambda = 0.4$\,mm is the screening length as proposed in Ref.~\cite{woerner2012}. The radial and vertical projections $F^{\text{rep}}_{\rho, z}$ of the Yukawa force computed from the experimental data are shown in Fig.~\ref{fig:9}. The figure also contains the best-rms fits introduced by the relations
\begin{equation}
\label{eq:fit}
F^{\text{rep}}_\rho = A_\rho \rho+B_\rho \rho^2, ~~~F^{\text{rep}}_z = A_z z,
\end{equation}
with $A_\rho, A_z, B_\rho$ the parameters and $\rho$ the cylindrical radius. In the case of counterclockwise rotation, the particles occasionally approach each other very closely (this may be an imaging issue), leading to a wide spread of the estimated Yukawa forces. Considering the large uncertainty of the parameters in Fig.~\ref{fig:9}\,(b), this approach may not be applicable there. 

% fig8  %%%%%%%%%%%%%%%%%%%%%%%%%%%%%%%%%%%%%%%%%%%%%%%%%%%%%%%%%%%%%%%%%%%%%%%
 %%%%%%%%%%%%%%%%%%%%%%%%%%%%%%%%%%%%%%%%%%%%%%%%%%%%%%%%%%%%%%%%%%%
%%%%%%%%%%%%%%%%%%%%%%%%%%%%%%%%%%%%%%%%%%%%%%%%%%%%%%%%%%%%%%%%%%%%%%%%%%%%%%%

% Neglect friction...
The friction force is $\mathbf{F}^{\text{fr}} = - M \gamma_{\text{Eps}} \mathbf{v}$, where $M = 1.1\times10^{-12}$~kg is the particle mass \cite{woerner2012}, $\gamma_{\text{Eps}}$ is the Epstein drag coefficient~\cite{epstein1924} and $\mathbf{v}$ is the particle velocity. The value of the friction force at the cluster periphery was estimated in Ref.~\cite{woerner2012} to be smaller than $10$\,fN. Compared to the Yukawa forces in the same region of the cluster, which are on the order of 0.5--2\,pN, the friction force is of minor importance and can be neglected in the balance of Eq.~(\ref{eq:balance}). 

%... centrifugal and Coriolis forces -> Calculate confinement
The inertial forces in the rotating frame of the cluster consist of the centrifugal force $\mathbf{F}^{\text{centri}}$ and the Coriolis force $\mathbf{F}^{\text{Cor}}$ \cite{hartmann2013}. The former is readily estimated as $F^{\text{centri}} = M \omega^2 \rho < 1$\,fN, where  $\omega \simeq 0.4$\,s$^{-1}$ is the rotation speed of the cluster. To estimate the particle velocities, five consecutive frames were averaged, as the jitter on some particles due to imaging processes complicates the calculation of the instantaneous velocity. The Coriolis force, which is maximal during transitions of a particle from an inner to an outer string or \textit{vice versa}, then yields $F^{\text{Cor}} = 2 M |\boldsymbol{\omega} \times \mathbf{v}| < 3$\,fN. Hence, compared to the Yukawa forces, the contribution of inertial forces can be neglected. The confinement force profiles can thus be estimated as ${F^{\text{conf}}_{\rho, z} \simeq - F^{\text{rep}}_{\rho, z}}$.

% fig9  %%%%%%%%%%%%%%%%%%%%%%%%%%%%%%%%%%%%%%%%%%%%%%%%%%%%%%%%%%%%%%%%%%%%%%%
 %%%%%%%%%%%%%%%%%%%%%%%%%%%%%%%%%%%%%%%%%%%%%%%%%%%%%%%%%%%%%%%%%%
%%%%%%%%%%%%%%%%%%%%%%%%%%%%%%%%%%%%%%%%%%%%%%%%%%%%%%%%%%%%%%%%%%%%%%%%%%%%%%%

% Get confinement frequencies.
Given the values of the fitting parameters~$A_{\rho}$ introduced above, the radial confinement near the center of the cluster can be estimated as
\begin{equation}
\label{eq:confinement_frequency}
F^{\text{conf}}_\rho \simeq - M \Omega_\rho^2 \rho, ~ M \Omega_\rho^2 \equiv A_\rho,
\end{equation}
where $\Omega_\rho$ is the cylindrical confinement parameter. The values of~$\Omega_\rho$ as well as the values of the vertical confinement parameter $\Omega_z$ (obtained similarly with the relation $M \Omega_z^2 \equiv A_z$) are shown in Table~\ref{table:1}.

% Vertical confinement stronger than radial?
It is not surprising that the vertical confinement force is systematically stronger than the radial one, as can be naturally expected due to the stronger forces of the sheath electric field and gravity compressing the cluster vertically~\cite{woerner2011}. Note that a stronger vertical confinement was also found for ``dust molecules'' consisting of only two particles~\cite{stokes2008}.

% Result
Neglecting the case of counterclockwise rotation due to the wide spread of data points, this approach yields a stronger radial confinement for the clockwise rotating cluster than for the nonrotating one (see Table~\ref{table:1}). This stronger confinement, increasing the \emph{cylindricity} of the rotating cluster, is in good agreement with our findings from the network analysis of Sec.~\ref{sec:networks_derived_from_distance_matrices}. %To study individual contributions to $\mathbf{F}^{\text{conf}}$ from, e.\,g., the ion drag force, dedicated measurements are needed. 

% Differences are not coming from rotation of individual particles.
From our estimate of the centrifugal and the Coriolis force it follows, furthermore, that the structural changes in the rotating cluster are not due to the cluster rotation \textit{per se}. It must rather be inferred that the electrostatic confinement of the whole cluster changes with the applied frequency, allowing for string formation within the rotating cluster.

% Dietmar
In Eq.~(\ref{eq:balance}), $\mathbf{F}^{\text{rep}}$ was assumed to consist of Yukawa interactions with constant particle charge $Ze$ and screening length~$\lambda$. It is straightforward to see that different values of~$\lambda$ and $Z$ only change the scale of the profiles in Fig.~\ref{fig:9}, and not the general shape. A dependency of the parameters on the vertical position of the particles, as can be expected in the sheath electric field, will change the force profiles, but one can guess that this will lead to the same result of a stronger radial confinement in the case of cluster rotation. While Yukawa interaction is valid and widely used in two-dimensional systems, an additional contribution from the wake-field interaction is not negligible in 3D~\cite{ivlev2012, lampe2000}. An adequate description of the wake field interaction is not straightforward and the subject of current research. 
Here, these anisotropic effects are included in the confinement force $\mathbf{F}^{\text{conf}}$, which is a mixture of external confinement and internal forces. The latter may be a significant contribution, as they result in the formation of particle strings.  Equation (~\ref{eq:fit}) is thus a fit to the global force profile, where the modulation due to the mutual wake-field interactions is not resolved. Nonetheless, our simplified model of the Yukawa-type interaction gives a hint for the origin of the differences in symmetry revealed by network analysis. A more detailed analysis should take into account the mass dispersion of the particles as well as the anisotropy in the interparticle interaction. 

% summary
\section{Summary}
\label{sec:summary}
To conclude, the data obtained with the help of a fully 3D holographic particle tracking  diagnostic \cite{kroll2008, woerner2012} allowed us to thoroughly explore the statistical as well as dynamical properties of particle clusters driven externally by rotating electric fields. The data was subjected to network analysis which demonstrated a significant difference between the well-studied case of nonrotating clusters and dynamically driven clusters. While the structuring is more spherical in the case of nonrotating clusters, the dynamically driven clusters have a more pronounced cylindrical structure. This difference is in agreement with the estimate of the radial confinement with the aid of a dynamical force balance. As for string detection, community assignment in networks proved a valuable tool for finding structures (without setting a predefined threshold) in systems that evolve over time.

\begin{acknowledgments}
The research leading to these results has received funding from the European Research Council under the European Union’s Seventh Framework Programme (FP7/2007-2013)/ERC Grant agreement 267499, and has been financed by DFG under grant No. SFB-TR24, project A3.
\end{acknowledgments}

% Literaturliste endgueltig anzeigen
\bibliography{./../../../10_papers/literature}

\begin{thebibliography}{41}
\expandafter\ifx\csname natexlab\endcsname\relax\def\natexlab#1{#1}\fi
\expandafter\ifx\csname bibnamefont\endcsname\relax
  \def\bibnamefont#1{#1}\fi
\expandafter\ifx\csname bibfnamefont\endcsname\relax
  \def\bibfnamefont#1{#1}\fi
\expandafter\ifx\csname citenamefont\endcsname\relax
  \def\citenamefont#1{#1}\fi
\expandafter\ifx\csname url\endcsname\relax
  \def\url#1{\texttt{#1}}\fi
\expandafter\ifx\csname urlprefix\endcsname\relax\def\urlprefix{URL }\fi
\providecommand{\bibinfo}[2]{#2}
\providecommand{\eprint}[2][]{\url{#2}}

\bibitem[{\citenamefont{Fortov et~al.}(2005)\citenamefont{Fortov, Ivlev,
  Khrapak, Khrapak, and Morfill}}]{fortov2005}
\bibinfo{author}{\bibfnamefont{V.~E.} \bibnamefont{Fortov}},
  \bibinfo{author}{\bibfnamefont{A.~V.} \bibnamefont{Ivlev}},
  \bibinfo{author}{\bibfnamefont{S.~A.} \bibnamefont{Khrapak}},
  \bibinfo{author}{\bibfnamefont{A.~G.} \bibnamefont{Khrapak}},
  \bibnamefont{and} \bibinfo{author}{\bibfnamefont{G.~E.}
  \bibnamefont{Morfill}}, \bibinfo{journal}{Phys. Rep.}
  \textbf{\bibinfo{volume}{421}}, \bibinfo{pages}{1} (\bibinfo{year}{2005}).

\bibitem[{\citenamefont{Ivlev et~al.}(2012)\citenamefont{Ivlev, L{\"o}wen,
  Morfill, and Royall}}]{ivlev2012}
\bibinfo{author}{\bibfnamefont{A.}~\bibnamefont{Ivlev}},
  \bibinfo{author}{\bibfnamefont{H.}~\bibnamefont{L{\"o}wen}},
  \bibinfo{author}{\bibfnamefont{G.}~\bibnamefont{Morfill}}, \bibnamefont{and}
  \bibinfo{author}{\bibfnamefont{C.~P.} \bibnamefont{Royall}},
  \emph{\bibinfo{title}{Complex Plasmas and Colloidal Dispersions:
  Particle-resolved Studies of Classical Liquids and Solids}}
  (\bibinfo{publisher}{World Scientific, Singapore}, \bibinfo{year}{2012}).

\bibitem[{\citenamefont{Arp et~al.}(2004)\citenamefont{Arp, Block, Piel, and
  Melzer}}]{arp2004}
\bibinfo{author}{\bibfnamefont{O.}~\bibnamefont{Arp}},
  \bibinfo{author}{\bibfnamefont{D.}~\bibnamefont{Block}},
  \bibinfo{author}{\bibfnamefont{A.}~\bibnamefont{Piel}}, \bibnamefont{and}
  \bibinfo{author}{\bibfnamefont{A.}~\bibnamefont{Melzer}},
  \bibinfo{journal}{Phys. Rev. Lett.} \textbf{\bibinfo{volume}{93}},
  \bibinfo{pages}{165004} (\bibinfo{year}{2004}).

\bibitem[{\citenamefont{Totsuji et~al.}(2005)\citenamefont{Totsuji, Ogawa,
  Totsuji, and Tsuruta}}]{totsuji2005}
\bibinfo{author}{\bibfnamefont{H.}~\bibnamefont{Totsuji}},
  \bibinfo{author}{\bibfnamefont{T.}~\bibnamefont{Ogawa}},
  \bibinfo{author}{\bibfnamefont{C.}~\bibnamefont{Totsuji}}, \bibnamefont{and}
  \bibinfo{author}{\bibfnamefont{K.}~\bibnamefont{Tsuruta}},
  \bibinfo{journal}{Phys. Rev. E} \textbf{\bibinfo{volume}{72}},
  \bibinfo{pages}{036406} (\bibinfo{year}{2005}).

\bibitem[{\citenamefont{Melzer et~al.}(2010)\citenamefont{Melzer,
  Buttensch{\"o}n, Miksch, Passvogel, Block, Arp, and Piel}}]{melzer2010}
\bibinfo{author}{\bibfnamefont{A.}~\bibnamefont{Melzer}},
  \bibinfo{author}{\bibfnamefont{B.}~\bibnamefont{Buttensch{\"o}n}},
  \bibinfo{author}{\bibfnamefont{T.}~\bibnamefont{Miksch}},
  \bibinfo{author}{\bibfnamefont{M.}~\bibnamefont{Passvogel}},
  \bibinfo{author}{\bibfnamefont{D.}~\bibnamefont{Block}},
  \bibinfo{author}{\bibfnamefont{O.}~\bibnamefont{Arp}}, \bibnamefont{and}
  \bibinfo{author}{\bibfnamefont{A.}~\bibnamefont{Piel}},
  \bibinfo{journal}{Plasma Phys. Controlled Fusion}
  \textbf{\bibinfo{volume}{52}}, \bibinfo{pages}{124028}
  (\bibinfo{year}{2010}).

\bibitem[{\citenamefont{W{\"o}rner et~al.}(2012)\citenamefont{W{\"o}rner,
  R{\"a}th, Nosenko, Zhdanov, Thomas, Morfill, Schablinski, and
  Block}}]{woerner2012}
\bibinfo{author}{\bibfnamefont{L.}~\bibnamefont{W{\"o}rner}},
  \bibinfo{author}{\bibfnamefont{C.}~\bibnamefont{R{\"a}th}},
  \bibinfo{author}{\bibfnamefont{V.}~\bibnamefont{Nosenko}},
  \bibinfo{author}{\bibfnamefont{S.~K.} \bibnamefont{Zhdanov}},
  \bibinfo{author}{\bibfnamefont{H.~M.} \bibnamefont{Thomas}},
  \bibinfo{author}{\bibfnamefont{G.~E.} \bibnamefont{Morfill}},
  \bibinfo{author}{\bibfnamefont{J.}~\bibnamefont{Schablinski}},
  \bibnamefont{and} \bibinfo{author}{\bibfnamefont{D.}~\bibnamefont{Block}},
  \bibinfo{journal}{Europhys. Lett.} \textbf{\bibinfo{volume}{100}},
  \bibinfo{pages}{35001} (\bibinfo{year}{2012}).

\bibitem[{\citenamefont{Hyde et~al.}(2013)\citenamefont{Hyde, Kong, and
  Matthews}}]{hyde2013}
\bibinfo{author}{\bibfnamefont{T.~W.} \bibnamefont{Hyde}},
  \bibinfo{author}{\bibfnamefont{J.}~\bibnamefont{Kong}}, \bibnamefont{and}
  \bibinfo{author}{\bibfnamefont{L.~S.} \bibnamefont{Matthews}},
  \bibinfo{journal}{Phys. Rev. E} \textbf{\bibinfo{volume}{87}},
  \bibinfo{pages}{053106} (\bibinfo{year}{2013}).

\bibitem[{\citenamefont{Arp et~al.}(2005)\citenamefont{Arp, Block, Klindworth,
  and Piel}}]{arp2005}
\bibinfo{author}{\bibfnamefont{O.}~\bibnamefont{Arp}},
  \bibinfo{author}{\bibfnamefont{D.}~\bibnamefont{Block}},
  \bibinfo{author}{\bibfnamefont{M.}~\bibnamefont{Klindworth}},
  \bibnamefont{and} \bibinfo{author}{\bibfnamefont{A.}~\bibnamefont{Piel}},
  \bibinfo{journal}{Phys. Plasmas} \textbf{\bibinfo{volume}{12}},
  \bibinfo{pages}{122102} (\bibinfo{year}{2005}).

\bibitem[{\citenamefont{Carstensen et~al.}(2010)\citenamefont{Carstensen,
  Greiner, and Piel}}]{carstensen2010}
\bibinfo{author}{\bibfnamefont{J.}~\bibnamefont{Carstensen}},
  \bibinfo{author}{\bibfnamefont{F.}~\bibnamefont{Greiner}}, \bibnamefont{and}
  \bibinfo{author}{\bibfnamefont{A.}~\bibnamefont{Piel}},
  \bibinfo{journal}{Phys. Plasmas} \textbf{\bibinfo{volume}{17}},
  \bibinfo{pages}{083703} (\bibinfo{year}{2010}).

\bibitem[{\citenamefont{Nosenko et~al.}(2009)\citenamefont{Nosenko, Ivlev,
  Zhdanov, Fink, and Morfill}}]{nosenko2009}
\bibinfo{author}{\bibfnamefont{V.}~\bibnamefont{Nosenko}},
  \bibinfo{author}{\bibfnamefont{A.~V.} \bibnamefont{Ivlev}},
  \bibinfo{author}{\bibfnamefont{S.~K.} \bibnamefont{Zhdanov}},
  \bibinfo{author}{\bibfnamefont{M.}~\bibnamefont{Fink}}, \bibnamefont{and}
  \bibinfo{author}{\bibfnamefont{G.~E.} \bibnamefont{Morfill}},
  \bibinfo{journal}{Phys. Plasmas} \textbf{\bibinfo{volume}{16}},
  \bibinfo{pages}{083708} (\bibinfo{year}{2009}).

\bibitem[{\citenamefont{Kong et~al.}(2011)\citenamefont{Kong, Hyde, Matthews,
  Qiao, Zhang, and Douglass}}]{kong2011}
\bibinfo{author}{\bibfnamefont{J.}~\bibnamefont{Kong}},
  \bibinfo{author}{\bibfnamefont{T.~W.} \bibnamefont{Hyde}},
  \bibinfo{author}{\bibfnamefont{L.}~\bibnamefont{Matthews}},
  \bibinfo{author}{\bibfnamefont{K.}~\bibnamefont{Qiao}},
  \bibinfo{author}{\bibfnamefont{Z.}~\bibnamefont{Zhang}}, \bibnamefont{and}
  \bibinfo{author}{\bibfnamefont{A.}~\bibnamefont{Douglass}},
  \bibinfo{journal}{Phys. Rev. E} \textbf{\bibinfo{volume}{84}},
  \bibinfo{pages}{016411} (\bibinfo{year}{2011}).

\bibitem[{\citenamefont{Melzer}(2006)}]{melzer2006}
\bibinfo{author}{\bibfnamefont{A.}~\bibnamefont{Melzer}},
  \bibinfo{journal}{Phys. Rev. E} \textbf{\bibinfo{volume}{73}},
  \bibinfo{pages}{056404} (\bibinfo{year}{2006}).

\bibitem[{\citenamefont{Tsytovich et~al.}(2007)\citenamefont{Tsytovich,
  Morfill, Fortov, Gusein-Zade, Klumov, and Vladimirov}}]{tsytovich2007}
\bibinfo{author}{\bibfnamefont{V.~N.} \bibnamefont{Tsytovich}},
  \bibinfo{author}{\bibfnamefont{G.~E.} \bibnamefont{Morfill}},
  \bibinfo{author}{\bibfnamefont{V.~E.} \bibnamefont{Fortov}},
  \bibinfo{author}{\bibfnamefont{N.~G.} \bibnamefont{Gusein-Zade}},
  \bibinfo{author}{\bibfnamefont{B.~A.} \bibnamefont{Klumov}},
  \bibnamefont{and} \bibinfo{author}{\bibfnamefont{S.~V.}
  \bibnamefont{Vladimirov}}, \bibinfo{journal}{New J. Phys.}
  \textbf{\bibinfo{volume}{9}}, \bibinfo{pages}{263} (\bibinfo{year}{2007}).

\bibitem[{\citenamefont{Ivanov and Melzer}(2009)}]{ivanov2009}
\bibinfo{author}{\bibfnamefont{Y.}~\bibnamefont{Ivanov}} \bibnamefont{and}
  \bibinfo{author}{\bibfnamefont{A.}~\bibnamefont{Melzer}},
  \bibinfo{journal}{Phys. Rev. E} \textbf{\bibinfo{volume}{79}},
  \bibinfo{pages}{036402} (\bibinfo{year}{2009}).

\bibitem[{\citenamefont{Pieper and Goree}(1996)}]{pieper1996dispersion}
\bibinfo{author}{\bibfnamefont{J.~B.} \bibnamefont{Pieper}} \bibnamefont{and}
  \bibinfo{author}{\bibfnamefont{J.}~\bibnamefont{Goree}},
  \bibinfo{journal}{Phys. Rev. Lett.} \textbf{\bibinfo{volume}{77}},
  \bibinfo{pages}{3137} (\bibinfo{year}{1996}).

\bibitem[{\citenamefont{Zuzic et~al.}(2000)\citenamefont{Zuzic, Ivlev, Goree,
  Morfill, Thomas, Rothermel, Konopka, S{\"u}tterlin, and
  Goldbeck}}]{zuzic2000}
\bibinfo{author}{\bibfnamefont{M.}~\bibnamefont{Zuzic}},
  \bibinfo{author}{\bibfnamefont{A.~V.} \bibnamefont{Ivlev}},
  \bibinfo{author}{\bibfnamefont{J.}~\bibnamefont{Goree}},
  \bibinfo{author}{\bibfnamefont{G.~E.} \bibnamefont{Morfill}},
  \bibinfo{author}{\bibfnamefont{H.~M.} \bibnamefont{Thomas}},
  \bibinfo{author}{\bibfnamefont{H.}~\bibnamefont{Rothermel}},
  \bibinfo{author}{\bibfnamefont{U.}~\bibnamefont{Konopka}},
  \bibinfo{author}{\bibfnamefont{R.}~\bibnamefont{S{\"u}tterlin}},
  \bibnamefont{and} \bibinfo{author}{\bibfnamefont{D.~D.}
  \bibnamefont{Goldbeck}}, \bibinfo{journal}{Phys. Rev. Lett.}
  \textbf{\bibinfo{volume}{85}}, \bibinfo{pages}{4064} (\bibinfo{year}{2000}).

\bibitem[{\citenamefont{Klumov et~al.}(2009)\citenamefont{Klumov, Huber,
  Vladimirov, Thomas, Ivlev, Morfill, Fortov, Lipaev, and
  Molotkov}}]{klumov2009}
\bibinfo{author}{\bibfnamefont{B.}~\bibnamefont{Klumov}},
  \bibinfo{author}{\bibfnamefont{P.}~\bibnamefont{Huber}},
  \bibinfo{author}{\bibfnamefont{S.}~\bibnamefont{Vladimirov}},
  \bibinfo{author}{\bibfnamefont{H.}~\bibnamefont{Thomas}},
  \bibinfo{author}{\bibfnamefont{A.}~\bibnamefont{Ivlev}},
  \bibinfo{author}{\bibfnamefont{G.}~\bibnamefont{Morfill}},
  \bibinfo{author}{\bibfnamefont{V.}~\bibnamefont{Fortov}},
  \bibinfo{author}{\bibfnamefont{A.}~\bibnamefont{Lipaev}}, \bibnamefont{and}
  \bibinfo{author}{\bibfnamefont{V.}~\bibnamefont{Molotkov}},
  \bibinfo{journal}{Plasma Phys. Controlled Fusion}
  \textbf{\bibinfo{volume}{51}}, \bibinfo{pages}{124028}
  (\bibinfo{year}{2009}).

\bibitem[{\citenamefont{Klumov}(2010)}]{klumov2010}
\bibinfo{author}{\bibfnamefont{B.~A.} \bibnamefont{Klumov}},
  \bibinfo{journal}{Physics-Uspekhi} \textbf{\bibinfo{volume}{53}},
  \bibinfo{pages}{1053} (\bibinfo{year}{2010}).

\bibitem[{\citenamefont{Antonova et~al.}(2006)\citenamefont{Antonova,
  Annaratone, Goldbeck, Yaroshenko, Thomas, and Morfill}}]{antonova2006}
\bibinfo{author}{\bibfnamefont{T.}~\bibnamefont{Antonova}},
  \bibinfo{author}{\bibfnamefont{B.~M.} \bibnamefont{Annaratone}},
  \bibinfo{author}{\bibfnamefont{D.~D.} \bibnamefont{Goldbeck}},
  \bibinfo{author}{\bibfnamefont{V.}~\bibnamefont{Yaroshenko}},
  \bibinfo{author}{\bibfnamefont{H.~M.} \bibnamefont{Thomas}},
  \bibnamefont{and} \bibinfo{author}{\bibfnamefont{G.~E.}
  \bibnamefont{Morfill}}, \bibinfo{journal}{Phys. Rev. Lett.}
  \textbf{\bibinfo{volume}{96}}, \bibinfo{pages}{115001}
  (\bibinfo{year}{2006}).

\bibitem[{\citenamefont{Lampe et~al.}(2000)\citenamefont{Lampe, Joyce, Ganguli,
  and Gavrishchaka}}]{lampe2000}
\bibinfo{author}{\bibfnamefont{M.}~\bibnamefont{Lampe}},
  \bibinfo{author}{\bibfnamefont{G.}~\bibnamefont{Joyce}},
  \bibinfo{author}{\bibfnamefont{G.}~\bibnamefont{Ganguli}}, \bibnamefont{and}
  \bibinfo{author}{\bibfnamefont{V.}~\bibnamefont{Gavrishchaka}},
  \bibinfo{journal}{Phys. Plasmas} \textbf{\bibinfo{volume}{7}},
  \bibinfo{pages}{3851} (\bibinfo{year}{2000}).

\bibitem[{\citenamefont{Lampe et~al.}(2005)\citenamefont{Lampe, Joyce, and
  Ganguli}}]{lampe2005}
\bibinfo{author}{\bibfnamefont{M.}~\bibnamefont{Lampe}},
  \bibinfo{author}{\bibfnamefont{G.}~\bibnamefont{Joyce}}, \bibnamefont{and}
  \bibinfo{author}{\bibfnamefont{G.}~\bibnamefont{Ganguli}},
  \bibinfo{journal}{IEEE Trans. Plasma Sci.} \textbf{\bibinfo{volume}{33}},
  \bibinfo{pages}{57} (\bibinfo{year}{2005}).

\bibitem[{\citenamefont{Watts and Strogatz}(1998)}]{watts1998}
\bibinfo{author}{\bibfnamefont{D.~J.} \bibnamefont{Watts}} \bibnamefont{and}
  \bibinfo{author}{\bibfnamefont{S.~H.} \bibnamefont{Strogatz}},
  \bibinfo{journal}{Nature (London)} \textbf{\bibinfo{volume}{393}},
  \bibinfo{pages}{440} (\bibinfo{year}{1998}).

\bibitem[{\citenamefont{Albert et~al.}(1999)\citenamefont{Albert, Jeong, and
  Barab{\'a}si}}]{albert1999}
\bibinfo{author}{\bibfnamefont{R.}~\bibnamefont{Albert}},
  \bibinfo{author}{\bibfnamefont{H.}~\bibnamefont{Jeong}}, \bibnamefont{and}
  \bibinfo{author}{\bibfnamefont{A.-L.} \bibnamefont{Barab{\'a}si}},
  \bibinfo{journal}{Nature (London)} \textbf{\bibinfo{volume}{401}},
  \bibinfo{pages}{130} (\bibinfo{year}{1999}).

\bibitem[{\citenamefont{Albert and Barab{\'a}si}(2002)}]{albert2002}
\bibinfo{author}{\bibfnamefont{R.}~\bibnamefont{Albert}} \bibnamefont{and}
  \bibinfo{author}{\bibfnamefont{A.-L.} \bibnamefont{Barab{\'a}si}},
  \bibinfo{journal}{Rev. Mod. Phys.} \textbf{\bibinfo{volume}{74}},
  \bibinfo{pages}{47} (\bibinfo{year}{2002}).

\bibitem[{\citenamefont{Barab{\'a}si}(2012)}]{barabasi2011}
\bibinfo{author}{\bibfnamefont{A.-L.} \bibnamefont{Barab{\'a}si}},
  \bibinfo{journal}{Nat. Phys.} \textbf{\bibinfo{volume}{8}},
  \bibinfo{pages}{14} (\bibinfo{year}{2012}).

\bibitem[{\citenamefont{Mucha et~al.}(2010)\citenamefont{Mucha, Richardson,
  Macon, Porter, and Onnela}}]{mucha2010}
\bibinfo{author}{\bibfnamefont{P.~J.} \bibnamefont{Mucha}},
  \bibinfo{author}{\bibfnamefont{T.}~\bibnamefont{Richardson}},
  \bibinfo{author}{\bibfnamefont{K.}~\bibnamefont{Macon}},
  \bibinfo{author}{\bibfnamefont{M.~A.} \bibnamefont{Porter}},
  \bibnamefont{and} \bibinfo{author}{\bibfnamefont{J.-P.}
  \bibnamefont{Onnela}}, \bibinfo{journal}{Science}
  \textbf{\bibinfo{volume}{328}}, \bibinfo{pages}{876} (\bibinfo{year}{2010}).

\bibitem[{\citenamefont{Kroll et~al.}(2008)\citenamefont{Kroll, Harms, Block,
  and Piel}}]{kroll2008}
\bibinfo{author}{\bibfnamefont{M.}~\bibnamefont{Kroll}},
  \bibinfo{author}{\bibfnamefont{S.}~\bibnamefont{Harms}},
  \bibinfo{author}{\bibfnamefont{D.}~\bibnamefont{Block}}, \bibnamefont{and}
  \bibinfo{author}{\bibfnamefont{A.}~\bibnamefont{Piel}},
  \bibinfo{journal}{Phys. Plasmas} \textbf{\bibinfo{volume}{15}},
  \bibinfo{pages}{063703} (\bibinfo{year}{2008}).

\bibitem[{\citenamefont{Lamb and Morales}(1983)}]{lamb1983}
\bibinfo{author}{\bibfnamefont{B.~M.} \bibnamefont{Lamb}} \bibnamefont{and}
  \bibinfo{author}{\bibfnamefont{G.~J.} \bibnamefont{Morales}},
  \bibinfo{journal}{Phys. Fluids} \textbf{\bibinfo{volume}{26}},
  \bibinfo{pages}{3488} (\bibinfo{year}{1983}).

\bibitem[{\citenamefont{Ludwig et~al.}(2012)\citenamefont{Ludwig, K{\"a}hlert,
  and Bonitz}}]{ludwig2012}
\bibinfo{author}{\bibfnamefont{P.}~\bibnamefont{Ludwig}},
  \bibinfo{author}{\bibfnamefont{H.}~\bibnamefont{K{\"a}hlert}},
  \bibnamefont{and} \bibinfo{author}{\bibfnamefont{M.}~\bibnamefont{Bonitz}},
  \bibinfo{journal}{Plasma Phys. Controlled Fusion}
  \textbf{\bibinfo{volume}{54}}, \bibinfo{pages}{045011}
  (\bibinfo{year}{2012}).

\bibitem[{\citenamefont{Donner et~al.}(2010)\citenamefont{Donner, Zou, Donges,
  Marwan, and Kurths}}]{donner2010}
\bibinfo{author}{\bibfnamefont{R.~V.} \bibnamefont{Donner}},
  \bibinfo{author}{\bibfnamefont{Y.}~\bibnamefont{Zou}},
  \bibinfo{author}{\bibfnamefont{J.~F.} \bibnamefont{Donges}},
  \bibinfo{author}{\bibfnamefont{N.}~\bibnamefont{Marwan}}, \bibnamefont{and}
  \bibinfo{author}{\bibfnamefont{J.}~\bibnamefont{Kurths}},
  \bibinfo{journal}{New J. Phys.} \textbf{\bibinfo{volume}{12}},
  \bibinfo{pages}{033025} (\bibinfo{year}{2010}).

\bibitem[{\citenamefont{Marwan et~al.}(2009)\citenamefont{Marwan, Donges, Zou,
  Donner, and Kurths}}]{marwan2009}
\bibinfo{author}{\bibfnamefont{N.}~\bibnamefont{Marwan}},
  \bibinfo{author}{\bibfnamefont{J.~F.} \bibnamefont{Donges}},
  \bibinfo{author}{\bibfnamefont{Y.}~\bibnamefont{Zou}},
  \bibinfo{author}{\bibfnamefont{R.~V.} \bibnamefont{Donner}},
  \bibnamefont{and} \bibinfo{author}{\bibfnamefont{J.}~\bibnamefont{Kurths}},
  \bibinfo{journal}{Phys. Lett. A} \textbf{\bibinfo{volume}{373}},
  \bibinfo{pages}{4246} (\bibinfo{year}{2009}).

\bibitem[{\citenamefont{Girvan and Newman}(2002)}]{girvan2002}
\bibinfo{author}{\bibfnamefont{M.}~\bibnamefont{Girvan}} \bibnamefont{and}
  \bibinfo{author}{\bibfnamefont{M.~E.~J.} \bibnamefont{Newman}},
  \bibinfo{journal}{Proc. Natl. Acad. Sci. U.S.A.}
  \textbf{\bibinfo{volume}{99}}, \bibinfo{pages}{7821} (\bibinfo{year}{2002}).

\bibitem[{\citenamefont{Newman and Girvan}(2004)}]{newman2004finding}
\bibinfo{author}{\bibfnamefont{M.~E.~J.} \bibnamefont{Newman}}
  \bibnamefont{and} \bibinfo{author}{\bibfnamefont{M.}~\bibnamefont{Girvan}},
  \bibinfo{journal}{Phys. Rev. E} \textbf{\bibinfo{volume}{69}},
  \bibinfo{pages}{026113} (\bibinfo{year}{2004}).

\bibitem[{\citenamefont{Newman}(2006)}]{newman2006finding}
\bibinfo{author}{\bibfnamefont{M.~E.~J.} \bibnamefont{Newman}},
  \bibinfo{journal}{Phys. Rev. E} \textbf{\bibinfo{volume}{74}},
  \bibinfo{pages}{036104} (\bibinfo{year}{2006}).

\bibitem[{\citenamefont{Bassett et~al.}(2013)\citenamefont{Bassett, Porter,
  Wymbs, Grafton, Carlson, and Mucha}}]{bassett2013}
\bibinfo{author}{\bibfnamefont{D.~S.} \bibnamefont{Bassett}},
  \bibinfo{author}{\bibfnamefont{M.~A.} \bibnamefont{Porter}},
  \bibinfo{author}{\bibfnamefont{N.~F.} \bibnamefont{Wymbs}},
  \bibinfo{author}{\bibfnamefont{S.~T.} \bibnamefont{Grafton}},
  \bibinfo{author}{\bibfnamefont{J.~M.} \bibnamefont{Carlson}},
  \bibnamefont{and} \bibinfo{author}{\bibfnamefont{P.~J.} \bibnamefont{Mucha}},
  \bibinfo{journal}{Chaos} \textbf{\bibinfo{volume}{23}},
  \bibinfo{pages}{013142} (\bibinfo{year}{2013}).

\bibitem[{\citenamefont{Blondel et~al.}(2008)\citenamefont{Blondel, Guillaume,
  Lambiotte, and Lefebvre}}]{blondel2008}
\bibinfo{author}{\bibfnamefont{V.~D.} \bibnamefont{Blondel}},
  \bibinfo{author}{\bibfnamefont{J.-L.} \bibnamefont{Guillaume}},
  \bibinfo{author}{\bibfnamefont{R.}~\bibnamefont{Lambiotte}},
  \bibnamefont{and} \bibinfo{author}{\bibfnamefont{E.}~\bibnamefont{Lefebvre}},
  \bibinfo{journal}{J. Stat. Mech. Theor. Exp.}
  \textbf{\bibinfo{volume}{2008}}, \bibinfo{pages}{P10008}
  (\bibinfo{year}{2008}).

\bibitem[{\citenamefont{Schella et~al.}(2013)\citenamefont{Schella, Mulsow,
  Melzer, Schablinski, and Block}}]{schella2013}
\bibinfo{author}{\bibfnamefont{A.}~\bibnamefont{Schella}},
  \bibinfo{author}{\bibfnamefont{M.}~\bibnamefont{Mulsow}},
  \bibinfo{author}{\bibfnamefont{A.}~\bibnamefont{Melzer}},
  \bibinfo{author}{\bibfnamefont{J.}~\bibnamefont{Schablinski}},
  \bibnamefont{and} \bibinfo{author}{\bibfnamefont{D.}~\bibnamefont{Block}},
  \bibinfo{journal}{Phys. Rev. E} \textbf{\bibinfo{volume}{87}},
  \bibinfo{pages}{063102} (\bibinfo{year}{2013}).

\bibitem[{\citenamefont{Epstein}(1924)}]{epstein1924}
\bibinfo{author}{\bibfnamefont{P.~S.} \bibnamefont{Epstein}},
  \bibinfo{journal}{Phys. Rev.} \textbf{\bibinfo{volume}{23}},
  \bibinfo{pages}{710} (\bibinfo{year}{1924}).

\bibitem[{\citenamefont{Hartmann et~al.}(2013)\citenamefont{Hartmann, Donk\'o,
  Ott, K\"ahlert, and Bonitz}}]{hartmann2013}
\bibinfo{author}{\bibfnamefont{P.}~\bibnamefont{Hartmann}},
  \bibinfo{author}{\bibfnamefont{Z.}~\bibnamefont{Donk\'o}},
  \bibinfo{author}{\bibfnamefont{T.}~\bibnamefont{Ott}},
  \bibinfo{author}{\bibfnamefont{H.}~\bibnamefont{K\"ahlert}},
  \bibnamefont{and} \bibinfo{author}{\bibfnamefont{M.}~\bibnamefont{Bonitz}},
  \bibinfo{journal}{Phys. Rev. Lett.} \textbf{\bibinfo{volume}{111}},
  \bibinfo{pages}{155002} (\bibinfo{year}{2013}).

\bibitem[{\citenamefont{W{\"o}rner et~al.}(2011)\citenamefont{W{\"o}rner,
  Nosenko, Ivlev, Zhdanov, Thomas, Morfill, Kroll, Schablinski, and
  Block}}]{woerner2011}
\bibinfo{author}{\bibfnamefont{L.}~\bibnamefont{W{\"o}rner}},
  \bibinfo{author}{\bibfnamefont{V.}~\bibnamefont{Nosenko}},
  \bibinfo{author}{\bibfnamefont{A.~V.} \bibnamefont{Ivlev}},
  \bibinfo{author}{\bibfnamefont{S.~K.} \bibnamefont{Zhdanov}},
  \bibinfo{author}{\bibfnamefont{H.~M.} \bibnamefont{Thomas}},
  \bibinfo{author}{\bibfnamefont{G.~E.} \bibnamefont{Morfill}},
  \bibinfo{author}{\bibfnamefont{M.}~\bibnamefont{Kroll}},
  \bibinfo{author}{\bibfnamefont{J.}~\bibnamefont{Schablinski}},
  \bibnamefont{and} \bibinfo{author}{\bibfnamefont{D.}~\bibnamefont{Block}},
  \bibinfo{journal}{Phys. Plasmas} \textbf{\bibinfo{volume}{18}},
  \bibinfo{pages}{063706} (\bibinfo{year}{2011}).

\bibitem[{\citenamefont{Stokes et~al.}(2008)\citenamefont{Stokes, Samarian, and
  Vladimirov}}]{stokes2008}
\bibinfo{author}{\bibfnamefont{J.~D.~E.} \bibnamefont{Stokes}},
  \bibinfo{author}{\bibfnamefont{A.~A.} \bibnamefont{Samarian}},
  \bibnamefont{and} \bibinfo{author}{\bibfnamefont{S.~V.}
  \bibnamefont{Vladimirov}}, \bibinfo{journal}{Phys. Rev. E}
  \textbf{\bibinfo{volume}{78}}, \bibinfo{pages}{036402}
  (\bibinfo{year}{2008}).

\end{thebibliography}

\end{document}